\documentclass[twocolumn,smallcondensed]{svjour3}

\usepackage[toc]{appendix}
\usepackage{epsfig}
\usepackage{amsfonts} 
\usepackage{amsmath}
\usepackage{color} 
\usepackage{hyperref}
\usepackage[square,sort,comma,numbers,sort&compress]{natbib} 
\usepackage{graphicx,amssymb}

\newcommand{\WK}{Wiener-Khinchin theorem }

\smartqed

\makeatletter
\expandafter\let\csname opt@amsmath.sty\endcsname\relax
\AtBeginDocument{
  \mathindent=15pt 
  \@mathmargin\@centering} 
\makeatother

\begin{document}

\title{$1/f^{\beta}$ noise  for scale-invariant processes: \\ 
How long you wait matters
}

\author{N. Leibovich   \and   E. Barkai
}
\institute{N. Leibovich \and E. Barkai \at
              Department of Physics, Institute of Nanotechnology and Advanced Materials,  Bar-Ilan University, Ramat-Gan 5290002, Israel  
}

\date{Received: date / Accepted: date}

\maketitle

\begin{abstract}
We study the power spectrum which is estimated from a nonstationary signal. In particular we examine the case when the signal is observed in a measurement time window $[t_w,t_w+t_m]$, namely the observation started after a waiting time $t_w$, and $t_m$ is the measurement duration. We introduce a generalized aging \WK which relates between the spectrum and the time- and ensemble-averaged correlation function for arbitrary $t_m$ and $t_w$. Furthermore we provide a general relation between the non-analytical behavior of the scale-invariant correlation function and the aging $1/f^{\beta}$ noise. We illustrate our general results with two-state renewal models with sojourn times' distributions having a broad tail.   
\end{abstract}

 \section{Introduction}
 The power spectrum of many natural processes exhibits  $1/f^{\beta}$ spectrum \cite{VanDerZiel,Dutta,Hooge81,Keshner,Weissman,Banerjee,Kaulakys,Yadav,Krapf,Rodriguez2015}. The spectrum is estimated from a recorded observable $I(t)$ which can be the light intensity, a spatial displacement, current, etc.  
Assume that such a signal $I(t)$ if found in several states in such a way that the sojourn times in some states are broadly distributed with fat tails \cite{bouchaud1992weak,lowen, Margolin04,Margolin05,Margolin06}.
The renewal assumption is usually invoked and thus the process is a continuous time random walk (CTRW) in the states space, where the number of states is fixed. Such a
system follows a power-law intermittency route to $1/f^{\beta}$ noise.
This means that power-law waiting times in a sub-state of the system are responsible for the observed $1/f^{\beta}$ spectrum. This approach was suggested as a fundamental mechanism for $1/f^{\beta}$
noise in the context of intermittency of chaos and turbulence in
the work of Manneville \cite{Manneville}, following the pioneering work of Mandelbrot \cite{Mandelbrot1967,graves2014brief,watkins2016mandelbrot}.
For example consider blinking-quantum dots, where under certain conditions, the emission of the dot switches from ``on'' to ``off'' state and vise-versa. The sojourn times in a state are broadly distributed in such a way that the average sojourn time diverges \cite{Margolin04,Margolin05,Margolin06,stefani2009beyond}. This well-investigated model yields $1/f^{\beta}$ noise,
 and as confirmed by recent experiments  \cite{Sadegh} the spectrum exhibits clear signs of nonstationarity: the power spectrum depends on the total  measurement time. 

The nonstationarity of a given process implies that we must consider carefully its initial conditions and the measurement protocol. For blinking-quantum dots the switching process from bright to dark  starts when the nano-system is bought into the spotlight of the exciting laser field. Namely one can  clearly identify when the process starts,
and this initiation is what we refer to as the origin of time $t=0$. Then the  sample spectrum is recorded between $t_w$ and $t_w + t_m$.
Here $t_w$ is the waiting time and $t_m$ is the time duration of the observation.  Another example is from the field of glassy dynamics \cite{Bouchaud,cugliandolo}.
There a quench of the system from say a high temperature is made, and then one waits time $t_w$ before recording some signal, say magnetization.  

The nonstationary route to $1/f^{\beta}$ started with the work of Mandelbrot \cite{Mandelbrot1967,watkins2016mandelbrot,graves2014brief}. Recently, the aged $1/f^{\beta}$ spectrum was found experimentally in the growing interface fluctuations in the $(1+1)$- dimensional KPZ class, using liquid-crystal turbulence \cite{Takeuchi2017}. This together with theoretical models, and the mentioned blinking quantum dots, motivate us to investigate the subject in further depth. 

For a nonstationary process one finds that, at least in principle, the measured observable depends both on $t_w$ and $t_m$. This means that whatever the observable is, whether it is the time-averaged position of a random walker \cite{schulz2013aging,schulz2014}, the power spectrum,  etc., we expect a dependence on both the waiting and the measurement times.  This is especially true for time-scale-invariant systems.
By this we mean a class of processes where the underlying ensemble-averaged  correlation function scales as $\langle I(t) I(t+\tau)\rangle  = t^\Upsilon \phi_{\rm EA}(\tau/t)$. Blinking quantum dot models,
renewal models, trap model, single-file diffusion, and many other processes fall into this wide category \cite{Margolin06,Leibovich,bouchaud1992weak,barrat,dechant2012,Kessler,taloni2010generalized}. In our recent works \cite{LeibovichPRL,LeibovichLong} we have found a simple,
yet general, relation between this scale-invariant correlation function and the measurement time-dependent spectrum. Our previous work considered
the case where the measurement starts at $t=0$, namely when $t_w=0$. Here we investigate the fingerprints  of the waiting time  $t_w\neq 0$ on the power spectrum, generalizing the \WK for this class of processes.
At the second stage of the article we demonstrate the theory for simple models of on-off blinking.

\section{The Waiting Time Dependent \WK}
\label{Main}

Here we derive a general formula for an aging spectrum where the measurement started at a certain time $t_w$ after time zero.  
In the following we generalize the previously published aging \WK \cite{LeibovichPRL,DechantPRL,LeibovichLong}.
Our goal is to introduce the relation between the measured spectral density to the time and ensemble- averaged correlation functions. 

\subsection{Aging \WK with the time-averaged correlation function}
\label{TA}

For a nonstationary process, the autocorrelation function is  $\left\langle I(t)I(t+\tau)\right\rangle=C(t,\tau)$, i.e. it depends on $t$ and the lag time $\tau$. $\langle.\rangle$ represents an ensemble average. We assume that the signal $I(t)$ starts evolving at time $t=0$. Then $I(t)$ is recorded in the time interval $(t_w,t_m+t_w)$, i.e. $t_w$ is a waiting time and $t_m$ is the measurement period. The sample spectrum is estimated through the periodogram
\begin{eqnarray}
&&\langle S_{t_w,t_m}(\omega)\rangle = 
\frac{1}{t_m}\left\langle \left| \int_{t_w}^{t_m+t_w}{\rm d}t I(t)e^{\imath\omega t}\right|^2 \right\rangle,
\label{eq:SampleSpectrumDefinition}
\end{eqnarray}
where the measurement time $t_m$ is long.
The spectrum thus is given by
\begin{eqnarray}
&&\langle S_{t_w,t_m}(\omega)\rangle = \label{eq:SampleSpectrum}\\ \nonumber
&&\frac{2}{t_m}\int_{t_w}^{t_m+t_w}{\rm d}t_1\int_{0}^{t_m+t_w-t_1}{\rm d}\tau \langle I(t_1)I(t_1+\tau) \rangle \cos(\omega\tau).
\end{eqnarray} 
We assume a scale invariant correlation function;
\begin{eqnarray}
\langle I(t)I(t+\tau) \rangle = t^{\Upsilon}\phi_{\rm EA}(\tau/t).
\label{eq:ScaleInveriantEA}
\end{eqnarray}
Here the sub-fix $(.)_{\rm EA}$ refers to the ensemble average. As was mentioned, this scaling condition is valid in many physical systems, e.g. see \cite{Margolin06,Leibovich,bouchaud1992weak,barrat,dechant2012,Kessler,taloni2010generalized}.
Here we assume that this scale invariance is valid for all $\tau$ and $t$. In reality this is an approximation which we discuss elsewhere \cite{LeibovichLong}, and briefly below. 
The time-averaged correlation function of the recorded signal is defined by
\begin{eqnarray}
C_{\rm TA}(t_w,t_m)=\frac{1}{t_m-\tau}\int_{t_w}^{t_w+t_m-\tau} {\rm d} t I(t) I(t+\tau).
\label{eq:TADefinition}
\end{eqnarray}
Taking the ensemble average over the time- averaged correlation function and using the scaled-invariant function Eq.~\eqref{eq:ScaleInveriantEA} gives the relation 
\begin{equation}
\langle C_{\rm TA}(t_m,t_w;\tau)\rangle =\frac{1}{t_m-\tau}\int_{t_w}^{t_m+t_w-\tau}{\rm d}t_1 t_1^{\Upsilon}
\phi_{\rm EA}\left(\frac{\tau}{t_1}\right),
\end{equation}
where the sub-fix $(.)_{\rm TA}$ refers to the time average. We thus find that the correlation function scales as
\begin{eqnarray}
\langle C_{\rm TA}(t_m,t_w;\tau)\rangle =t_m^{\Upsilon}\varphi_{\rm TA}\left(\frac{\tau}{t_m};\frac{t_w}{t_m}\right),
\label{eq:ScaledTA}
\end{eqnarray}
where the relation between $\phi_{\rm EA}(x)$ and $\varphi_{\rm TA}(x,{\cal T})$ is 
\begin{eqnarray}
\varphi_{\rm TA}\left(x;{\cal T}\right) = \frac{x^{\Upsilon+1}}{1-x}\int_{x/(1-{\cal T}-x)}^{x/{\cal T}} {\rm d}x \frac{\phi_{\rm EA}(x)}{x^{2+\Upsilon}},
\label{eq:TArelationEA}
\end{eqnarray}
where $x=\tau/t_m$ and ${\cal T}=t_w/t_m$. 
Changing the integration order in Eq.~\eqref{eq:SampleSpectrum} using Eq.~\eqref{eq:ScaledTA} gives
\begin{equation}
\langle S_{t_w,t_m}(\omega)\rangle =2t_m^{\Upsilon+1} \int_0^1 {\rm d}x(1-x) \varphi_{\rm TA}(x,{\cal T})\cos(\tilde{\omega}x),
\label{eq:TimeAverage}
\end{equation}
where $\tilde{\omega}=\omega t_m$. This gives the relation between the time-averaged autocorrelation function and the spectrum. 
For a further detailed derivation see App.~\ref{app:derivation}.

\subsection{Aging \WK with the ensemble-averaged correlation function}
\label{EA}

In this subsection, we derive a relation between the ensemble- averaged
autocorrelation function $C(t,\tau)=t^{\Upsilon}\phi_{\rm EA}(\tau/t) $ and the time-dependent power spectrum. 
For simplicity we first assume that $\Upsilon=0$. We obtain
\begin{eqnarray}
&&\langle S_{t_w,t_m}(\omega)\rangle = 2t_m \int_0^{1/{\cal T}}{\rm d}y\phi_{\rm EA}\left(y \right) \label{eq:09 }\nonumber \\
 && \left\lbrace \frac{\cos\left[\tilde{\omega}(1+{\cal T})\frac{y}{1+y} \right]}{\tilde{\omega}^2y^2} +\frac{\sin\left[\tilde{\omega}(1+{\cal T})\frac{y}{1+y} \right](1+{\cal T})}{\tilde{\omega}y(1+y)} \right. \nonumber \\ 
&&\left.
 - \frac{\cos\left[\tilde{\omega}{\cal T}y \right]}{\tilde{\omega}^2y^2} -\frac{\sin\left[\tilde{\omega}{\cal T}y \right]{\cal T}}{\tilde{\omega}y}
\right\rbrace,
\end{eqnarray}
see derivations in App.~\ref{app:derivation}. The relation between the ensemble-averaged correlation function and the spectrum for $\Upsilon \neq 0$ is given by
\begin{eqnarray}
\langle S_{t_w,t_m}(\omega)\rangle && = 2t_m^{\Upsilon+1} \int_0^{1/{\cal T}}{\rm d}y\phi_{\rm EA}\left(y \right) \nonumber \\
 && \Re \left\lbrace \frac{(-\imath)^{\Upsilon}\Gamma \left [2+\Upsilon,-\imath\tilde{\omega}(1+{\cal T})\frac{y}{1+y}\right]}{(\tilde{\omega}y)^{2+\Upsilon}}  \right. \nonumber \\ 
&&\left.
 -\frac{(-\imath)^{\Upsilon}\Gamma \left [2+\Upsilon,\imath\tilde{\omega}{\cal T}y \right]}{(\tilde{\omega}y)^{2+\Upsilon}}  
\right\rbrace 
\label{eq:EnsembleAverage}
\end{eqnarray}
where  $\Gamma(a,x)=\int_x^{\infty}{\rm d}t t^{\alpha-1}e^{-t}$ refers to the incomplete Gamma function and $\Re[.]$ represents the real part. In the limit of ${\cal T}\ll 1$ (i.e. $t_w \ll t_m$) we recover our previous published results for both the time averaged formalism , Eq.~\eqref{eq:TimeAverage}, and the ensemble averaged formalism, Eq.~\eqref{eq:EnsembleAverage}, see  \cite{LeibovichPRL,LeibovichLong}. 
We note that both aging Wiener-Khinchin relations, Eqs.~\eqref{eq:TimeAverage} and \eqref{eq:EnsembleAverage} are equivalent.
 
\subsection{The aging $1/f^{\beta}$ noise}
\label{1/f}

In the following we show that when $\phi_{\rm EA}(x)$ is a non-analytic function in the vicinity of zero the spectrum is of $1/f^{\beta}$ type. Consider ensemble-averaged correlation function with 
\begin{eqnarray}
C(t,\tau)=t^{\Upsilon}\phi_{\rm EA}\left(\frac{\tau}{t}\right )\approx t^{\Upsilon}\left[A-B \left (\frac{\tau}{t}\right)^{V}\right] 
\label{eq:CorrelationScale}
\end{eqnarray}
in the limit of $\tau \ll t$. Here $A$ and $B$ are constants which determined by the specific process. We demand $0<|V|<1$ and $\Upsilon-V>-1$ for convergence.  In the limit of long time and $\omega t_m \gg 1$ we find
\begin{eqnarray}
\langle S_{t_w,t_m}(\omega)\rangle \approx \tilde{B}_V \Lambda_{\Upsilon-V+1}\left(\frac{t_w}{t_m}\right)t_m^{\Upsilon-V}\omega^{-1-V}
\label{eq:1/f_waiting}
\end{eqnarray}
where the aging factor is 
\begin{eqnarray}
\Lambda_{\nu}(x)=\left[\left(x+1\right)^{\nu}-x^{\nu}\right]
\end{eqnarray}
 with $\nu>0$, and  $\tilde{B}_V=2 B\sin(\pi V/2)\Gamma(1+V)/(1~-~V+~\Upsilon)$.  Here we conclude that the spectrum depends on  both measurement time $t_m$ and waiting time $t_w$. The dependence of $t_w$ and $t_m$ is a direct outcome of an aged process.
 Considering our previously published results for $t_w=0$ \cite{LeibovichPRL,LeibovichLong}
 we further obtain from Eq.~\eqref{eq:1/f_waiting} the relation 
\begin{eqnarray}
\langle S_{t_w,t_m}(\omega)\rangle ~=~\Lambda_{\Upsilon-V+1}\left(\frac{t_w}{t_m}\right) \langle S_{t_w=0,t_m}(\omega)\rangle.
\label{eq:GeneralRelation}
\end{eqnarray}

A non-analytic correlation function, Eq.~\eqref{eq:CorrelationScale}, is found in many processes (see for example the table given in \cite{LeibovichLong}). 
In that sense our general result Eqs.~\eqref{eq:1/f_waiting},\eqref{eq:GeneralRelation} are universal.  The aging prefactor $\Lambda_{\nu}(x)$ is also found  for other observables (beyond the power spectrum) in some  nonstationary processes, in particular the CTRW and in models of deterministic  intermittency \cite{schulz2014,akimoto2013aging}. For example the aged and non-aged time averaged mean-square displacement (MSD) fulfills similar relation. See, for example, the continuous time random walk (CTRW)  \cite{schulz2014,Niemann2016}, heterogeneous diffusion processes  \cite{cherstvy2015ergodicity},  and scaled Brownian motion  \cite{safdari2015quantifying}. Therefore the function $\Lambda_{\nu}(x)$  appears rather naturally for several observables
and, as we have shown here, it is the outcome of the scale invariance of the correlation function and thus not limited to CTRWs. 
We comment, though, that the exponent $\nu$, which is defined trough $\Lambda_{\nu}(x)$, is bounded $0<\nu<1$  in the CTRW  \cite{Niemann2016,schulz2014}.  Here, such a constrain on the upper bound is not necessary, where in principle the value of $\nu$ may by equal or larger than~1. 

When $\Upsilon=V$ we find $\langle S_{t_w,t_m}(\omega)\rangle \approx \tilde{B}_V  \omega^{-1-V}$, since $\Lambda_{1}\left(x\right)=1$ for every $x$. It means that the $1/f^{\beta}$ noise seems stationary, i.e. it neither depends on the measurement time nor the waiting time. Nevertheless, the appearance of the time-independent $1/f^{\beta}$ noise does not mean that the underlining process is stationary. See for example the displacement of a tracer particle in a single-file diffusion model \cite{LeibovichLong}. Further distinction must be made with respect to bounded and non-bounded processes. For a bounded process  whose variance is asymptotically non-zero, i.e. $\Upsilon=0$, e.g. the blinking quantum dot model in Sec.~\ref{Infinite} below, we get when $t_w = 0$, the behavior predicted by Mandelbrot, i.e. $S_t(\omega)\sim t^{\alpha-1}\omega^ {\alpha-2}$, which as explained in  \cite{Mandelbrot1967,Niemann}  solves the famous low-frequency-cutoff paradox of the $1/f^{\beta}$ noise.

We examine the behavior of the $1/f^{\beta}$ spectrum for two cases; a slightly aging spectrum; when $ t_w \ll t_m$ and  strong aging when $t_m \ll t_w$.   In the limit $t_w\ll t_m$ the system ``forgets'' its initial states and we find
\begin{eqnarray}
\langle S_{t_m}(\omega)\rangle \approx \tilde{B}_V t_m^{\Upsilon-V}\omega^{-1-V},
\label{eq:1/f_waiting_tm}
\end{eqnarray}
which recovers previous results \cite{LeibovichPRL,LeibovichLong}. In the opposite limit $t_w\gg t_m$ we obtain
\begin{eqnarray}
\langle S_{t_w}(\omega)\rangle \approx \tilde{B}_V (\Upsilon-V+1) t_w^{\Upsilon-V}\omega^{-1-V},
\label{eq:1/f_waiting_tw}
\end{eqnarray}
then the spectrum depends only on the waiting time.

We note that for positive value of $V$, i.e. $0<V<1$, the term $(-B)t^{\Upsilon-V} \tau^{V}$ in Eq.~\eqref{eq:CorrelationScale} is the second leading order (e.g. blinking quantum dot with infinite mean sojourn times in Sec.~\ref{Infinite} ), while for negative $V$, i.e. when $-1<V<0$, the leading term is $(-B)t^{\Upsilon-V}\tau^{V}$ where $B$ must be negative (e.g. blinking quantum dot with finite mean ``on'' times, see Sec.~\ref{finite}). For both cases, negative and positive $V$, Eqs.~\eqref{eq:1/f_waiting}, \eqref{eq:1/f_waiting_tm} and \eqref{eq:1/f_waiting_tw} are valid.  

\section{Renewal Models for Blinking Quantum Dots}
\label{BlinkingQD}
We consider a simple renewal model with a two-state system, where $I(t)=0$ is the state ``off'' and $I(t)=I_0$ is ``on''. Without lost of generality we choose the system to be initially at $I(0)=I_0$. At random times $t_n$ the system switches to the other state alternately (``on'' $\rightarrow$ ``off'' or ``off'' $\rightarrow$ ``on''). The renewal times are $t_n=\sum_1^n\tau_i$ where the  sojourn time in a substate are given by the sequence $\left\{\tau_i\right\}_{i=1}^n$. The integer $n$ represents the number of renewals until time $t$.  The waiting times in ``off'' state  are independent identically distributed with common probability density function (PDF) $\psi_{\rm off}(\tau)\propto \tau^{-1-\alpha}$ where $0<\alpha<1$. Hence,  the mean waiting time in a substate ``off'' diverges. In our models the distribution of the ``on'' sojourn times $\psi_{\rm on}(\tau)$ may have infinite or finite mean. 
Therefore we consider two cases; in the first one the ``on'' and ``off'' times are fat-tailed distributed with $\psi_{\rm off}(\tau)=\psi_{\rm on}(\tau)$. In the second case we consider ``on'' times with a finite mean distribution while ``off'' times are distributed with power law as in the first case. We note that both cases were examined experimentally before \cite{Pelton,Sadegh}

As was mentioned in the introduction this stochastic process is used to analyzed the blinking quantum dot process \cite{Margolin04,Margolin06,Margolin05}, or  turbulent flows \cite{herault2015experimental}  (though there $\alpha>1$ so we do not consider this case in detail).
Further, the renewal model is closely related to the trap model in glasses  (where the exponent $\alpha$ is related to the ratio between the temperature and the glass transition temperature) \cite{Bouchaud} and the L\'evy walk model, where the walker's velocity is renewed \cite{zaburdaev2015levy}.

\subsection{Infinite Mean ``On'' Sojourn Time Distribution}
\label{Infinite}
Here we consider that both ``on'' and ``off'' times are power-law distributed
\begin{eqnarray}
\psi_{\rm off/on}(\tau)\sim (\tau_0/\tau)^{1+\alpha} \ \ \ \ \ \ \ \ \ \tau>\tau_0
\label{eq:LongTailDistribution}
\end{eqnarray}
 where $\tau_0$ is a microscopic time scale and $0<\alpha<1$ (see for example the experiment in \cite{Pelton}). We choose for both substates, ``on'' and ``off'', the same exponent $\alpha$ for simplicity.
Certain aspects of this model were studied analytically before \cite{lowen,Margolin04,Margolin05,Margolin06,Niemann,godreche,Lukovic,aquino2011transmission}.

\begin{figure}
\includegraphics[width=1.05\columnwidth,trim=25 0 0 10]{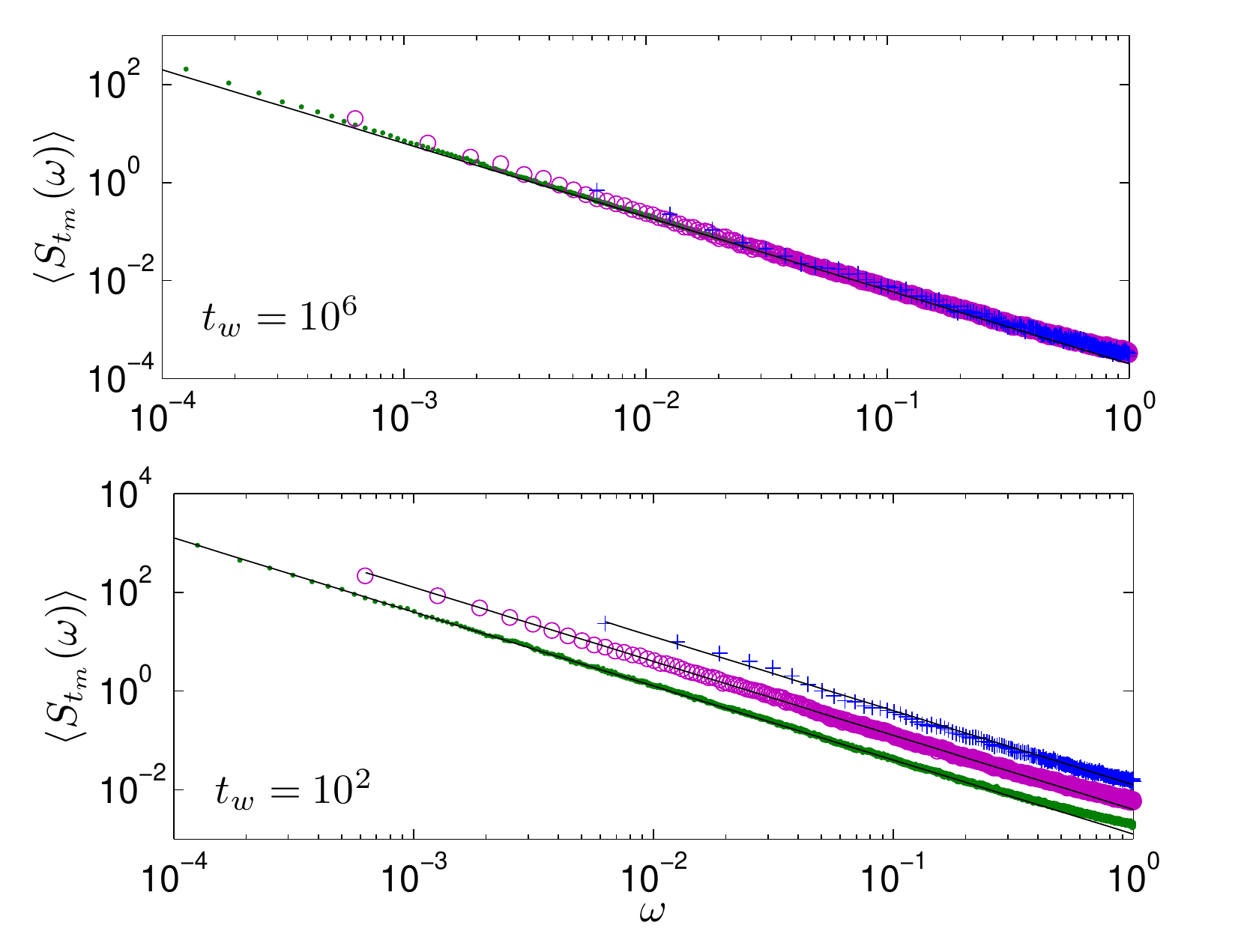}
		\caption{Simulation results for the spectrum recorded in the time interval $[t_w,t_w+t_m]$ from the quantum dot renewal model with infinite mean ``on'' and ``off'' times. Here we use $I_0=1$, $\alpha=0.5$ and $10^4$ realizations for the ensemble average. The measurement times are $t_m=10^3$ (blue crosses), $t_m=10^4$ (pink circles), and $t_m=10^5$ (green dots). In the upper panel we present the results for a fixed waiting time $t_w=10^6$ which is greater than $t_m$, and in the lower panel $t_w=10^2$ which is shorter than $t_m$.  The solid lines represent the analytic result Eq.~\eqref{eq:20}. The spectrum is given at natural frequencies; $\omega=2\pi n/t_m$ where $n\in {\mathbb{N}}$. The lowest recorded frequency is $2\pi/t_m$ hence the spectrum is shifted to the red.   	}
	\label{fig:WaitingBQD}
\end{figure}

The analytic formula for the ensemble-averaged correlation function is given in \cite{Margolin06,godreche}, when $t,\tau \gg \tau_0$,
\begin{equation}
\langle I(t)I(t+\tau)\rangle =I_0^2 \left[\frac{1}{2}-\frac{1}{4}\frac{\sin(\pi\alpha)}{\pi}B\left(\frac{\tau}{\tau+t},1-\alpha,\alpha \right)\right],
\label{eq:BetaCorrelationBQD}
\end{equation}
where $B(x,a,b)$ is the incomplete Beta function.
In the limit of $\tau \ll t$ we find
\begin{eqnarray}
C(t,\tau)\approx I_0^2\left[\frac{1}{2}-\frac{\sin(\pi\alpha)}{4(1-\alpha)\pi}\left(\frac{\tau}{t} \right)^{1-\alpha}\right],
\end{eqnarray}
so $\Upsilon=0$ and $V=1-\alpha$. Therefore using equation \eqref{eq:1/f_waiting}, we obtain in the limit of $\omega t_m\gg 1$ the aging $1/f^{\beta}$ noise, where $\beta=2-\alpha$;
\begin{equation}
\langle  S_{t_w,t_m}(\omega)\rangle \approx I_0^2 \frac{\cos(\pi\alpha/2)}{2\Gamma(1+\alpha)}\Lambda_{\alpha}\left(\frac{t_w}{t_m}\right) t^{\alpha-1}\omega^{\alpha-2}.
\label{eq:20}
\end{equation}
For this example we recover the result given in \cite{Niemann2016}. There the renewal model was used, while here we derive the results trough general arguments using the autocorrelation function properties. In the limit of short waiting time, $t_w\ll t_m$, the system ``forgets'' its initial state and we find 
\begin{eqnarray}
\langle  S_{t_m}(\omega)\rangle \approx I_0^2 \frac{\cos(\pi\alpha/2)}{2\Gamma(1+\alpha)}t_m^{\alpha-1}\omega^{\alpha-2},
\label{eq:21}
\end{eqnarray}
while in the opposite limit, where $t_w\gg t_m$ we find
\begin{eqnarray}
\langle  S_{t_w}(\omega)\rangle \approx I_0^2 \frac{\alpha \cos(\pi\alpha/2)}{2\Gamma(1+\alpha)}t_w^{\alpha-1}\omega^{\alpha-2}.
\label{eq:22}
\end{eqnarray}
In Fig.~\ref{fig:WaitingBQD} we show the simulation results (symbols) with $\alpha=0.5$, fixed waiting time $t_w=10^6$ (upper panel) and $t_w=10^2$ (lower panel) and three measurement times; $t_m=10^3$, $t_m=10^4$, and $t_m=10^5$. The analytic predictions present a good agreement with the simulation. 
We mention that the nonstationary spectrum, and in particular the dependence of the spectrum on the waiting time $t_w$, Eq.~\eqref{eq:20}, has been found in glassy dynamics \cite{Bouchaud}.

\subsection{Finite Mean ``On'' Sojourn Time Distribution}
\label{finite}
\begin{figure}
\centering
\includegraphics[width=1.05\columnwidth,trim=25 0 0 0]{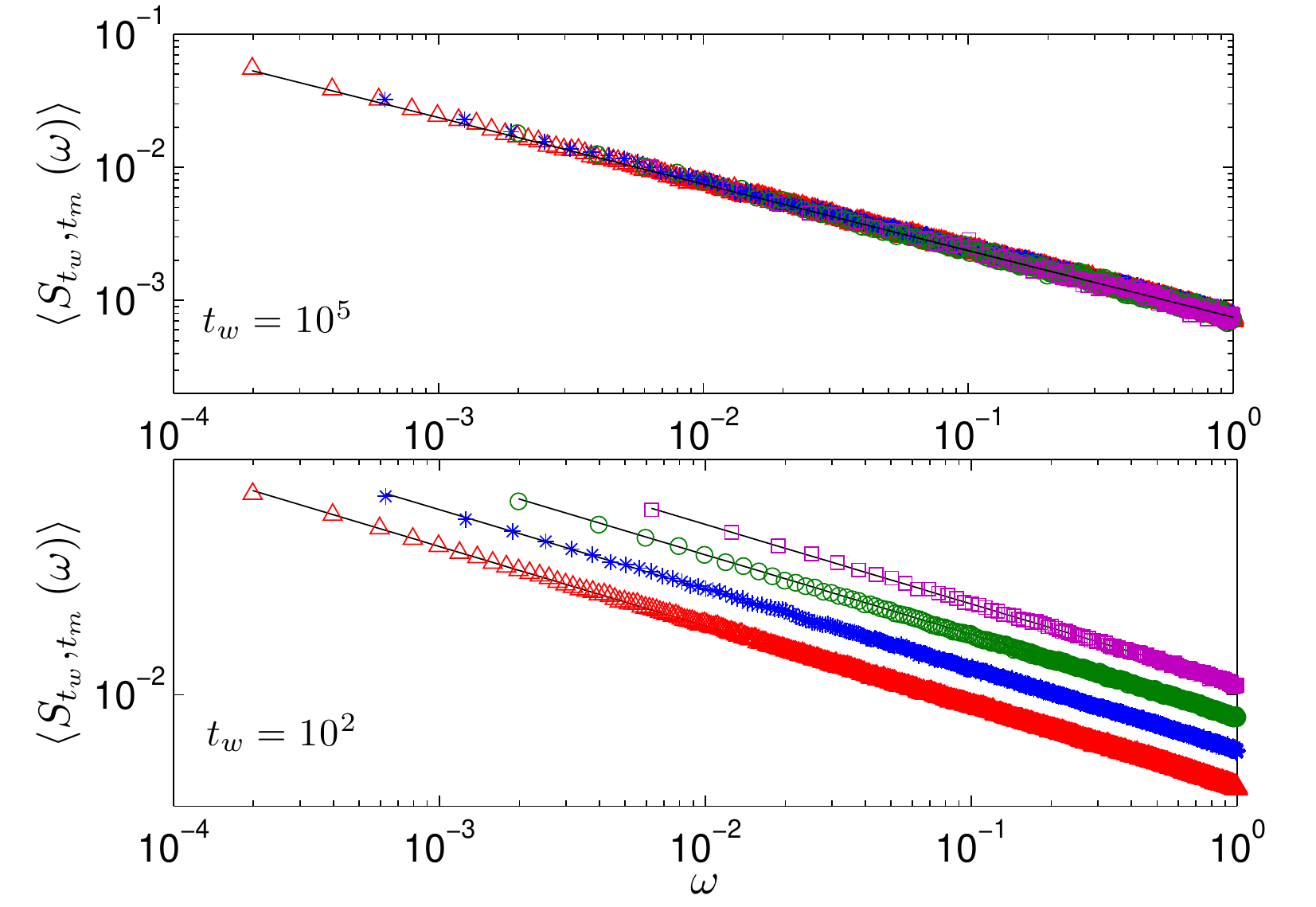}
\caption{Simulation of the spectrum for the renewal model with finite mean ``on'' sojourn time $\langle \tau \rangle =1$, and heavy tailed distributed ``off'' time with $\alpha=0.5$. Here we use $10^4$ realizations for the ensemble average. The waiting time are fixed, $t_w=10^5$ (upper panel) and $t_w=10^2$ (lower panel). The measurement times are $t_m=10^3$ (pink squares), $t_m=3162$ (green circles), $t_m=10^4$ (blue stars), and $t_m=31622$ (red triangles). The analytic prediction (solid lines) Eq.~\eqref{eq:waitingBQDfin} presents a nice agreement with the simulation when $\omega t_m=2\pi n \gg 1$, where $n$ is an integer.     }
\label{fig:WaitingBQDFinite}
\end{figure}

Now we consider that the PDFs for the ``on'' and ``off'' sojourn time are given in Laplace space by
\begin{eqnarray}
\psi_{\rm on}(s)&\approx& 1- \langle \tau\rangle s + ... \nonumber\\ 
\psi_{\rm off}(s)&\approx& 1- a s^{\alpha} + ...
\label{eq:PsiOn}
\end{eqnarray}
for small $s$, $a=\tau_0^{\alpha}\Gamma(1-\alpha)$ and $0<\alpha<1$. It refers to the case where the ``on'' times have finite mean $\langle\tau\rangle$, for example they are drawn from
an exponential distribution,  and the ``off'' times are fat-tailed distributed as before.
The correlation function is given by \cite{Margolin04}
\begin{eqnarray}
C(t,\tau)\approx t^{2\alpha-2} \left[I_0^2 \frac{\langle \tau \rangle ^2 }{a^2\Gamma^2 (\alpha)} \left(\frac{\tau}{t} \right)^{\alpha-1} \right],
\label{eq:CorrelationFinite}
\end{eqnarray} 
where both $t$ and $\tau$ are assumed to be long.
Here we find that $\Upsilon=2\alpha-2$, $V=\alpha-1$, $A=0$ and $B=-I_0^2 \langle \tau \rangle ^2 /[a^2\Gamma^2 (\alpha)]$. 
Using Eq.~\eqref{eq:1/f_waiting}, we obtain 
\begin{eqnarray}
\langle  S_{t_w,t_m}(\omega)\rangle \approx I_0^2 \frac{2\langle\tau\rangle ^2 \cos(\pi\alpha/2)}{a^2\Gamma(1+\alpha)}\Lambda_{\alpha}\left(\frac{t_w}{t_m}\right) t_m^{\alpha-1}\omega^{-\alpha},
 \nonumber \\
\ \label{eq:waitingBQDfin}
\end{eqnarray}
in the limit of $\omega t_m \gg 1$. Hence in the limit $t_m \gg t_w$ we obtain
\begin{eqnarray}
\langle S _{t_m}(\omega)\rangle \approx I_0^2\frac{2\langle \tau\rangle ^2 \cos(\pi\alpha/2)}{\Gamma(\alpha) a^2 \alpha }t_m^{\alpha-1}\omega^{-\alpha}.
\end{eqnarray}
When $t_w\gg t_m$ we find
\begin{eqnarray}
\langle S _{t_w}(\omega)\rangle \approx I_0^2\frac{2\langle \tau\rangle ^2 \cos(\pi\alpha/2)}{\Gamma(\alpha) a^2  }t_w^{\alpha-1}\omega^{-\alpha}
\end{eqnarray}
which depends on the waiting time and independent of the measurement time $t_m$. 
In Fig.~\ref{fig:WaitingBQDFinite} we present the simulation results for the process where we use $I_0=~1$, the ``on'' sojourn times are exponentially distributed with  $\langle \tau \rangle=1$, $\alpha=0.5$, and $\tau_0=1$. The waiting time $t_w$ are fixed; in the upper panel $t_w=10^5$ and in the lower panel $t_w=10^2$. The simulations nicely agree with the analytic prediction Eq.~\eqref{eq:waitingBQDfin} when $\omega t_m =2\pi n \gg 1$ for $n\in {\mathbb{N}}$.   

\section{ Continuous Frequencies Spectra }
\begin{figure}
\includegraphics[width=1.1\columnwidth, trim= 30 0 0 10]{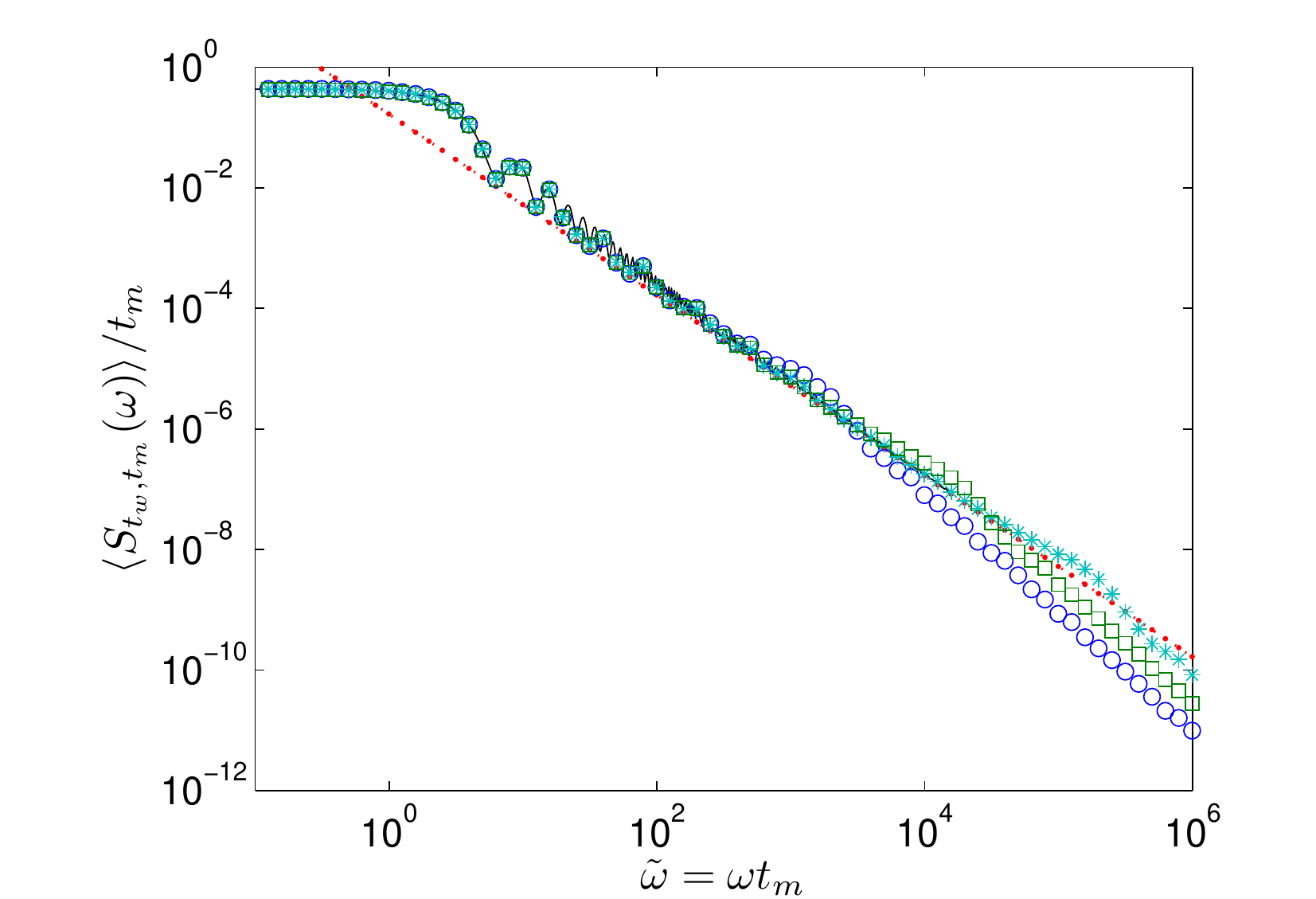}
		\caption{Simulation results (symbols) for the spectrum recorded in the time interval $[t_w,t_w+t_m]$ from a quantum dot renewal model with infinite average ``on'' and ``off'' times. Here we use $I_0=1$, $\alpha=0.5$, and $10^4$ realizations for ensemble averaging. In addition we fixed ${\cal T}=1$ and present the results for $t_m=t_w=10^3$ (blue circles), $t_m=t_w=10^4$ (green squares), and $t_m=t_w=10^5$ (cyan stars). A numerical estimation based on the analytic relation Eq.~\eqref{eq:09 } using the exact correlation function Eq.~\eqref{eq:BetaCorrelationBQD} is presented by black line. The analytic prediction Eq.~\eqref{eq:20} is presented with red dotted line. On the natural frequencies $\tilde{\omega} = 2 \pi n$ the results fall on a nearly straight line as shown in Fig.~\ref{fig:WaitingBQD}. 	}
	\label{fig:WaitingBQDContin}
\end{figure}

\begin{figure}
\centering
\includegraphics[width=1.1\columnwidth, trim= 20 0 0 0]{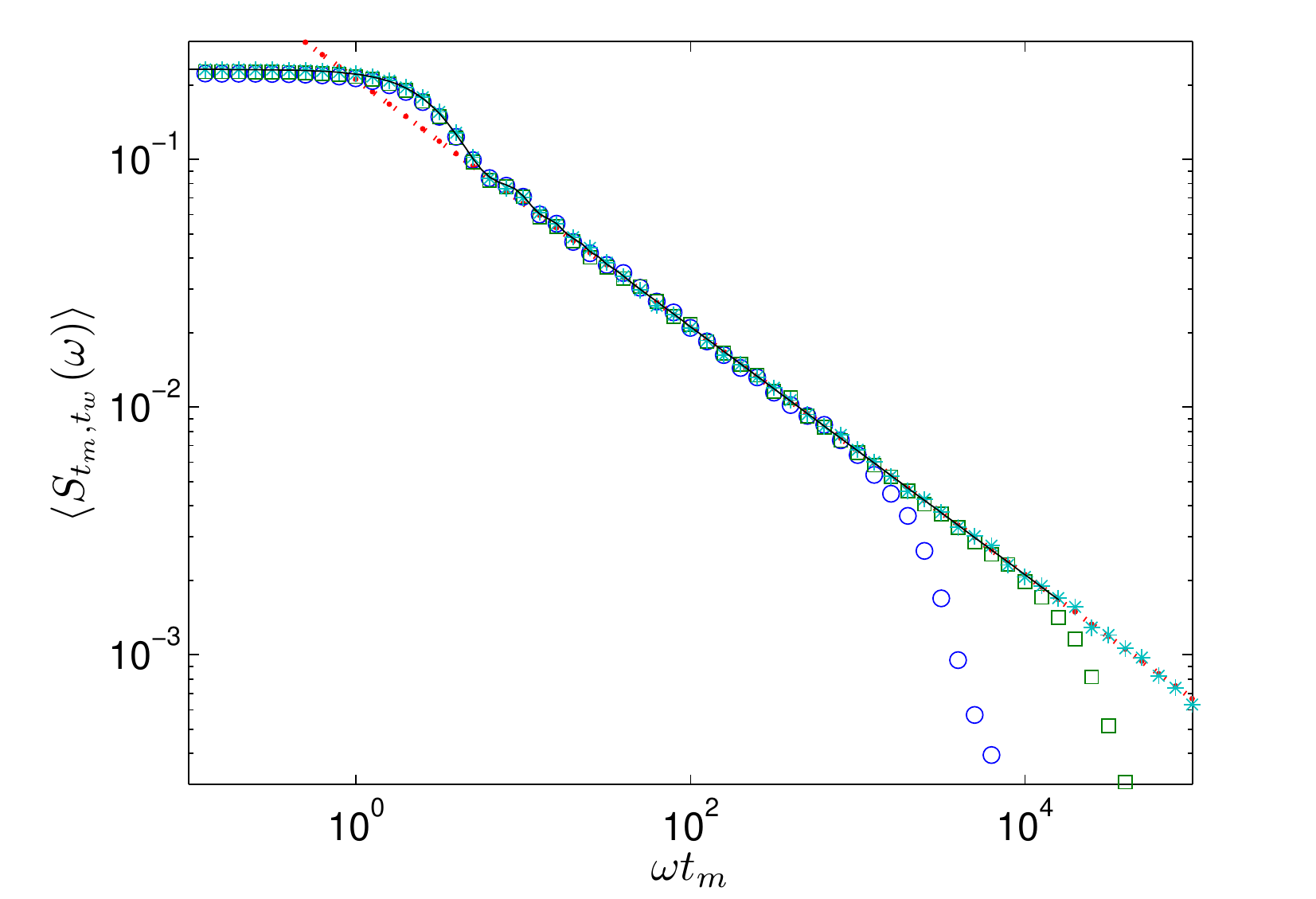}
\caption{Simulation of the spectrum for the renewal model with finite mean ``on'' sojourn time $\tau_0=\langle \tau \rangle =1$, and heavy tailed distributed ``off'' time with $\alpha=0.5$. Here we use $10^4$ realizations for the ensemble average. The ratio between the waiting time and the measurement time is fixed; ${\cal T}=1$. The simulation are presented for $t_w=t_m=10^3$ (blue circles), $t_w=t_m=10^4$ (green squares), and $t_w=t_m=10^5$ (cyan stars). A numerical integration of the exact results Eq.~\eqref{eq:EnsembleAverage} is presented by black line. The red dotted line represents the analytic prediction Eq.~\eqref{eq:waitingBQDfin}.  }
\label{fig:WaitingFinite}
\end{figure}

 In Figs.~\ref{fig:WaitingBQD} and \ref{fig:WaitingBQDFinite} we presented the $1/f^{\beta}$ spectrum showing its time dependence. On a log-log plot these curves are, with a good approximation, a straight line.  The $1/f^{\beta}$ spectrum was found in the limit $ \omega t_m \gg 1$. This means that when $\omega t_m= 2\pi$ (for example) we can find small deviations from $1/f^{\beta}$ noise. These deviations  are difficult to detect (see figures \ref{fig:WaitingBQD}, \ref{fig:WaitingBQDFinite}). It should be noted that the aging Wiener Khinchin theorem provides full information on the correlation function (and vice versa) namely the $1/f^{\beta}$ spectrum contains only a partial information on the shape of the correlation function for its small arguments, see Eqs.~\eqref{eq:CorrelationScale} and \eqref{eq:1/f_waiting}. Figs.~\ref{fig:WaitingBQD},\ref{fig:WaitingBQDFinite} are plotted on a physically natural scale, namely $\omega= 2 \pi n /t_m$ where $n\in {\mathbb{N}}$, as is the standard choice in the analysis of noise \cite{kubo2012statistical}. Plotting the power spectrum using continuous frequencies (this is certainly easy to do with a computer) reveals a richer structure, see figs.~\ref{fig:WaitingBQDContin} and \ref{fig:WaitingFinite}. These spectra reveal oscillations which are an effect of the finite measurement time. For further discussion limited for $t_w=0$ see \cite{LeibovichLong}.


The aging \WK depends on the correlation functions in their scaling form. In Eqs.~\eqref{eq:TimeAverage}, \eqref{eq:09 }, \eqref{eq:EnsembleAverage} we assume that such a scale invariance is valid for all times, however Eqs.~\eqref{eq:BetaCorrelationBQD}, \eqref{eq:waitingBQDfin} are valid for long times. We find, though, that detailed information on the waiting time PDFs $\psi(\tau)$ both in the ``on'' and ``off'' states are not crucial, besides a few variables (like $\alpha$). When one samples very large frequencies and finite measurement time the spectrum will depend on the fine details of the model (when $t_m$ and $t_w$ are fixed), since the scaling form of the correlation functions breaks in the short time limit. Therefore we obtain deviations from asymptotic analytic predictions at very high frequencies, see Figs.~\ref{fig:WaitingBQDContin} and \ref{fig:WaitingFinite}. The 
order of taking limits of frequency and time is thus importance.
We will discuss this issue in details in a future publication.

 In Figs.~\ref{fig:WaitingBQDContin},\ref{fig:WaitingFinite} we compare between  simulations and the numerical estimation of the analytic relations Eqs.~\eqref{eq:09 },  \eqref{eq:EnsembleAverage} using the exact correlation function Eqs.~\eqref{eq:BetaCorrelationBQD},  \eqref{eq:CorrelationFinite}. For the numerics we use the standard numerical integration of Mathematica, see details in App.~\ref{app:numerical}. We see a good agreement with theory, and as we increase both $t_m$ and $t_w$ (with fixed ${\cal T}$) the agreement becomes better. 

\section{Summary}
We have shown that the sample spectrum, which is estimated from a nonstationary process, is affected by both the waiting time and the measurement time. 
We introduce new formulas relating the time and ensemble average correlation functions to the sample spectrum. Further we show that a non-analytic correlation function provides $1/f^{\beta}$ noise, with a universal aging prefactor $\Lambda_{\nu}(t_w/t_m)$.  
These general predictions were tested successfully for two-state processes. While the theory relies on scale-invariant correlation functions, these are valid only for long times. Simulations show that convergence to asymptotic results are easily reached. 

\begin{acknowledgements}
EB thanks Jean-Philippe Bouchaud,   Holger Kantz, Diego Krapf, Kazumasa Takeuchi and Nick Watkins for insights, collaborations, and discussions. 
\end{acknowledgements}

\appendix
\section{Detailed derivations of the Results in Sec.~2}
\label{app:derivation}
Here we provide the detailed derivation of the results given in Sec.~\ref{Main}. Similar derivation for the specific case  $t_w=0$ can be found in \cite{LeibovichLong}. Our starting point is the sample spectrum given in Eq.~\eqref{eq:SampleSpectrum};
\begin{eqnarray}
&&\langle S_{t_w,t_m}(\omega)\rangle =\label{eq:29}\\ \nonumber
&&\frac{2}{t_m}\int_{t_w}^{t_m+t_w}{\rm d}t_1\int_{0}^{t_m+t_w-t_1}{\rm d}\tau \langle I(t_1)I(t_1+\tau) \rangle \cos(\omega\tau).
\end{eqnarray}
Change the integration order
\begin{eqnarray}
&&\langle S_{t_w,t_m}(\omega)\rangle =\\ \nonumber
&&\frac{2}{t_m}\int_{0}^{t_m}{\rm d}\tau\cos(\omega\tau) \int_{t_w}^{t_m+t_w-\tau}{\rm d}t_1 \langle I(t_1)I(t_1+\tau) \rangle,
\label{eq:30}
\end{eqnarray}
and substitute Eq.~\eqref{eq:TADefinition} gives
\begin{eqnarray}
&&\langle S_{t_w,t_m}(\omega)\rangle =\frac{2}{t_m}\int_{0}^{t_m}{\rm d}\tau \cos(\omega\tau) (t_m-\tau)\langle C_{\rm TA}(t_w,t_m;\tau) \rangle \nonumber \\
&&
=2t_m \int_{0}^{1}{\rm d}x \cos(\omega t_m  x) (1-x)\langle C_{\rm TA}(t_w,t_m;x) \rangle.
\label{eq:31}
\end{eqnarray}
We use the scale-invariant correlation function; $\langle I(t)I(t+~\tau) \rangle =t^{\Upsilon}\phi_{\rm EA}(\tau /t)$ and find
\begin{eqnarray}
\langle C_{\rm TA}(t_w,t_m;\tau) \rangle &&=\frac{1}{t_m-\tau}\int_{t_w}^{t_m+t_w-\tau}{\rm d}t \langle I(t)I(t+\tau) \rangle  \nonumber\\
&&= \frac{1}{t_m-\tau}\int_{t_w}^{t_m+t_w-\tau}{\rm d}t t^{\Upsilon}\phi_{\rm EA}(\tau/t).
\end{eqnarray}
Scaling the variables $x=\tau/t_m$, ${\cal T}=t_w/t_m$ and $\tilde{t}=t/t_m$ gives
\begin{eqnarray}
\langle C_{\rm TA}(t_w,t_m;x) \rangle &&=\frac{t_m^{\Upsilon}}{1-x}\int_{{\cal T}}^{1+{\cal T}-x}{\rm d}{\tilde{t}} \tilde{t}^{\Upsilon} \phi_{\rm EA}(x/\tilde{t}).
\end{eqnarray}
Then we change the integration variable $y=x/\tilde{t}$ and find
\begin{eqnarray}
\langle C_{\rm TA}(t_w,t_m;x) \rangle &&=t_m^{\Upsilon}\frac{x^{1+\Upsilon}}{1-x}\int_{x/(1+{\cal T}-x)}^{x/{\cal T}}{\rm d} y \frac{\phi_{\rm EA}(y)}{y^{\Upsilon+2}} \nonumber \\ 
&&= t_m^{\Upsilon} \varphi_{\rm TA}(x,{\cal T}),
\label{eq:34}
\end{eqnarray} 
thus Eq.~\eqref{eq:TArelationEA} is recovered. Substitute Eq.~\eqref{eq:34} into Eq.~\eqref{eq:31} recovers Eq.~\eqref{eq:TimeAverage} in the text.

For the derivation of Eqs.~\eqref{eq:09 } and \eqref{eq:EnsembleAverage} we start with Eq.~\eqref{eq:29} and substitute the scale invariant correlation function 
\begin{eqnarray}
&&\langle S_{t_w,t_m}(\omega)\rangle =\\ \nonumber
&&\frac{2}{t_m}\int_{t_w}^{t_m+t_w}{\rm d}t_1 t_1^{\Upsilon} \int_{0}^{t_m+t_w-t_1}{\rm d}\tau \phi_{\rm EA} (\tau/t_1) \cos(\omega\tau).
\end{eqnarray}
After scaling to dimensionless integration variables; $\tilde{t_1}=t_1/t_m$ and $\tilde{\tau}=\tau/t_m$, we obtain
\begin{eqnarray}
&&\langle S_{t_w,t_m}(\omega)\rangle =\\ \nonumber
&&2t_m^{1+\Upsilon}\int_{\cal T}^{1+{\cal T}}{\rm d}\tilde{t_1} \tilde{t_1}^{\Upsilon} \int_{0}^{1+{\cal T}-\tilde{t_1}}{\rm d}\tilde{\tau}  \phi_{\rm EA}(\tilde{\tau}/\tilde{t_1}) \cos(\omega t_m \tilde{\tau}).
\end{eqnarray}
We change the integration variable of the inner integral $x=\tilde{\tau}/\tilde{t_1}$ and find
\begin{eqnarray}
&&\langle S_{t_w,t_m}(\omega)\rangle =\\ \nonumber
&&2t_m^{1+\Upsilon}\int_{\cal T}^{1+{\cal T}}{\rm d}\tilde{t_1}\int_{0}^{(1+{\cal T}-\tilde{t_1})/\tilde{t_1}}{\rm d}x \tilde{t_1}^{\Upsilon+1} \phi_{\rm EA}(x) \cos(\tilde{\omega} x \tilde{t_1}),
\end{eqnarray}
where $\tilde{\omega}=\omega t_m$. Then we swap the integration order, i.e. 
\begin{eqnarray}
&&\langle S_{t_w,t_m}(\omega)\rangle =\\ \nonumber
&&2t_m^{1+\Upsilon}\int_{0}^{1/{\cal T}}{\rm d}x \phi_{\rm EA}(x) \int_{\cal T}^{(1+{\cal T})/(1+x)}{\rm d}\tilde{t_1} \tilde{t_1}^{\Upsilon+1} \cos(\tilde{\omega} x \tilde{t_1}),
\end{eqnarray}
solve the inner integral and recover Eqs.~\eqref{eq:09 } and \eqref{eq:EnsembleAverage}.
We note that both  $\phi_{\rm EA}(x)$ and $\varphi_{\rm TA}(x)$ have units of $[{\rm Intensity}^2 {\rm time}^{-\Upsilon}]$. Therefore the power spectrum has units of $[{\rm Intensity}^2 {\rm time}]$ as it should be, since it presents the measured power for a given frequency.
 
\section{Numerical estimation of Eqs.~(9) and (10) and Simulation Details }
\label{app:numerical}

 We choose $\alpha=1/2$ for both models. This specific value simplifies the numerical estimation for the following reason. For the infinite mean ``on'' time we find the correlation function given in Eq.~\eqref{eq:BetaCorrelationBQD} is given by
 \begin{eqnarray}
 C(t,\tau)\approx \frac{I_0^2}{2} \left[1-\frac{1}{\pi} \arcsin \left(\sqrt{\frac{\tau}{\tau+t}} \right) \right].
 \end{eqnarray}
Substituting in Eq.~\eqref{eq:09 } and using standard numerical integrator in Mathematica  gives the results which are presented in Fig.~\ref{fig:WaitingBQDContin}.

For the finite ``on'' time case, we find $\Upsilon=-1$ (since $\alpha=1/2$) therefore using  Eq.~\eqref{eq:EnsembleAverage} we obtain
\begin{eqnarray}
&&\langle S_{t_m,t_w}(\omega)\rangle = \\ &&2\int_0^{1/{\cal T}}{\rm d}y \phi_{\rm EA}(y) \left\{   \frac{\sin \left(\frac{(1+{\cal T})\tilde{\omega}y}{1+y}\right)}{\tilde{\omega} y} -\frac{\sin\left({\cal T}\tilde{\omega y}\right) }{\tilde{\omega }y}\right\} \nonumber.
\end{eqnarray}
Therefore, using Eq.~\eqref{eq:CorrelationFinite} for $\alpha=1/2$, we obtain
\begin{eqnarray}
&&\langle S_{t_w,t_m}(\omega) \rangle = I_0^2\frac{2 \langle \tau\rangle ^2}{\pi ^2 \tau_0 } \\ \nonumber
&&\int_{0}^{1/{\cal T}}{\rm d}y
y^{-1/2}\left\{   \frac{\sin \left(\frac{(1+{\cal T})\tilde{\omega}y}{1+y}\right)}{\tilde{\omega} y} -\frac{\sin\left({\cal T}\tilde{\omega y}\right) }{\tilde{\omega }y}\right\}.
\end{eqnarray}
Numerical integration using Mathematica provides the results given in Fig.~\ref{fig:WaitingFinite}.

For simulation we use the periodogram Eq.~\eqref{eq:SampleSpectrumDefinition}. Without lost of generality we use $I(t=0)=I_0$. We use the simulation method given in the appendix of \cite{LeibovichLong}. There only generating the sojourn times sequence $\{ \tau_n \}$ is needed. For generating randomly heavy-tailed distributed sojourn times we use $\tau=(1-U)^{-1/\alpha}$ where $U\in [0,1]$ is uniformly distributed, and $\alpha=1/2$. The distribution of $\tau$ is hence $\psi(\tau)=\alpha \tau^{-1-\alpha}$ for $\tau\geq 1$ (here $\tau_0=1$). The exponential distributed times with average $\langle\tau\rangle$ are given with $\tau=-\log(U)\langle \tau \rangle$, where we use $\langle\tau\rangle=~1$.

 \bibliographystyle{spphys}
 \bibliography{bib1} 
 
\end{document}